\documentclass[aps,prl,twocolumn,showpacs]{revtex4-2}

\usepackage{amssymb}
\usepackage{amsmath}
\usepackage{graphicx}
\usepackage{color}

\begin{document}
\title{Photon correlation Fourier spectroscopy of a B center in hBN}
\author{A. Delteil, S. Buil, J.-P. Hermier}

\affiliation{ Universit\'e Paris-Saclay, UVSQ, CNRS,  GEMaC, 78000, Versailles, France.
}

\begin{abstract}
The potential of solid-state quantum emitters for applications critically depends on several key figures of merit. One of the most important is the coherence time of the emitted single photons, which can be compromised by fast dephasing and spectral diffusion. In hexagonal boron nitride (hBN), blue-emitting color centers (or B centers) are seen as favorable in this regard, in the light of prior studies mainly based on resonant excitation. Yet, their time-coherence properties in the more accessible regime of non-resonant excitation (or photoluminescence) have not been extensively characterized. Here, we investigate the coherence and spectral diffusion of the photoluminescence from a B center in the continuous wave regime using photon correlation Fourier spectroscopy. We determine that the emission lineshape consists of a homogeneous contribution, whose linewidth increases with the laser power, and which is broadened by spectral diffusion at a timescale of 10 to 100~$\mu$s. At low power and short time, the emission line is only a factor $\sim$2 above the Fourier limit, while at long times, the inhomogeneous linewidth increases up to more than a gigahertz. Our work deepens the understanding of decoherence processes of this preeminent family of quantum emitters in hBN.
\end{abstract}


 \maketitle
\section{I. Introduction}

Single-photon emitters (SPEs) in the solid-state are essential building blocks for photonic quantum technologies~\cite{Aharonovich16, Pelucchi22, Couteau23}. Hexagonal boron nitride (hBN) as a host material offers many favorable aspects for these applications, such as the possiblity to be manipulated and integrated down to the single layer scale without constraints of lattice matching. The high quality of the light-emitting defects, or color centers, that can be obtained by various methods in this two-dimensional material triggers a growing number of applications to quantum optics and quantum technologies~\cite{Cacan25}. Among these quantum emitters, the blue-emitting SPEs, or B centers, can be locally created with an electron beam~\cite{Fournier21, Gale22, Roux22} and have been shown to be bright and stable, and exhibit narrow linewidths at low temperature~\cite{Fournier21, Horder22}.

The suitability of quantum emitters for applications is strongly determined by the coherence time of the emitted photons. For SPEs embedded in solid-state matrices, the main detrimental processes typically include pure dephasing -- which reflects intrinsic, rapid loss of coherence due to short-timescale interactions (\textit{e.g.}, phonons) --, and spectral diffusion (SD), which arises from slower environmental fluctuations that cause time-dependent shifts of the emission frequency. These two processes call for different mitigation strategies: pure dephasing can be alleviated through Purcell enhancement in photonic structures, whereas SD requires stabilization of the local environment, for instance via electrical contacts, or improved crystal quality. It is therefore crucial to discriminate between and quantitatively characterize these contributions when benchmarking a quantum emitter. Another impacting parameter is the excitation regime, which plays a major role in the impact that the environment has on the figures of merit of solid-state emitters~\cite{Monniello14, Nawrath19, Scholl19}. Photoluminescence (PL) is based on non-resonant laser excitation, which typically degrades the coherence of SPEs by introducing excess noise due to high energy charges and heat generated by a high laser power. These effects can be mitigated by using a resonant laser that only addresses the transition of interest.

In the case B centers, in PL, the indistinguishability between consecutive photons is limited to about 0.5~\cite{Fournier23PRA} -- however, a full coherence is restored at short timescales under resonant laser drive~\cite{Fournier23PRB, Gerard25, Gerard25arxiv}. The associated linewidth then remains close to the Fourier limit during tens to hundreds of microseconds, before being broadened by spectral diffusion (SD) that randomizes the emission wavelength within a $\sim$GHz wide envelope~\cite{Horder25, Gerard25}. Despite these benefits, the resonant regime makes collection of the zero-phonon-line (ZPL) emission at the exact same wavelength as the excitation laser particularly challenging~\cite{Gerard25arxiv}. By contrast, photoluminescence greatly facilitates the filtering of the excitation laser from the ZPL emission. An in-depth identification and characterization of dephasing and SD in this excitation regime is therefore essential.

Standard dispersive (frequency-domain) spectroscopy lacks both the necessary time and spectral resolution to infer the coherence and SD characteristics of narrow-line quantum emitters undergoing fast SD. While Fourier-transform spectroscopy (FTS) has a better spectral resolution, it also requires averaging over macroscopic times, preventing to measure fast SD~\cite{Marshall11}. In both cases, the time resolution exceeds the inverse count rate, which yields a typical cutoff of at least a few milliseconds. On the other hand, photon correlation Fourier spectroscopy (PCFS) allows to overcome this limit by relying on photon coincidences from the output ports of an interferometer, which allows to accumulate information about the fast dynamics down to the resolution of single-photon detectors~\cite{Brokmann06, Coolen07}. Therefore, this technique has the potential to identify the various contributions to the line broadening of solid-state emitters, and makes it possible to characterize their fast dynamics and their dependence on external parameters.

\begin{figure*}[t]
  \centering
  \includegraphics[width=0.95\linewidth]{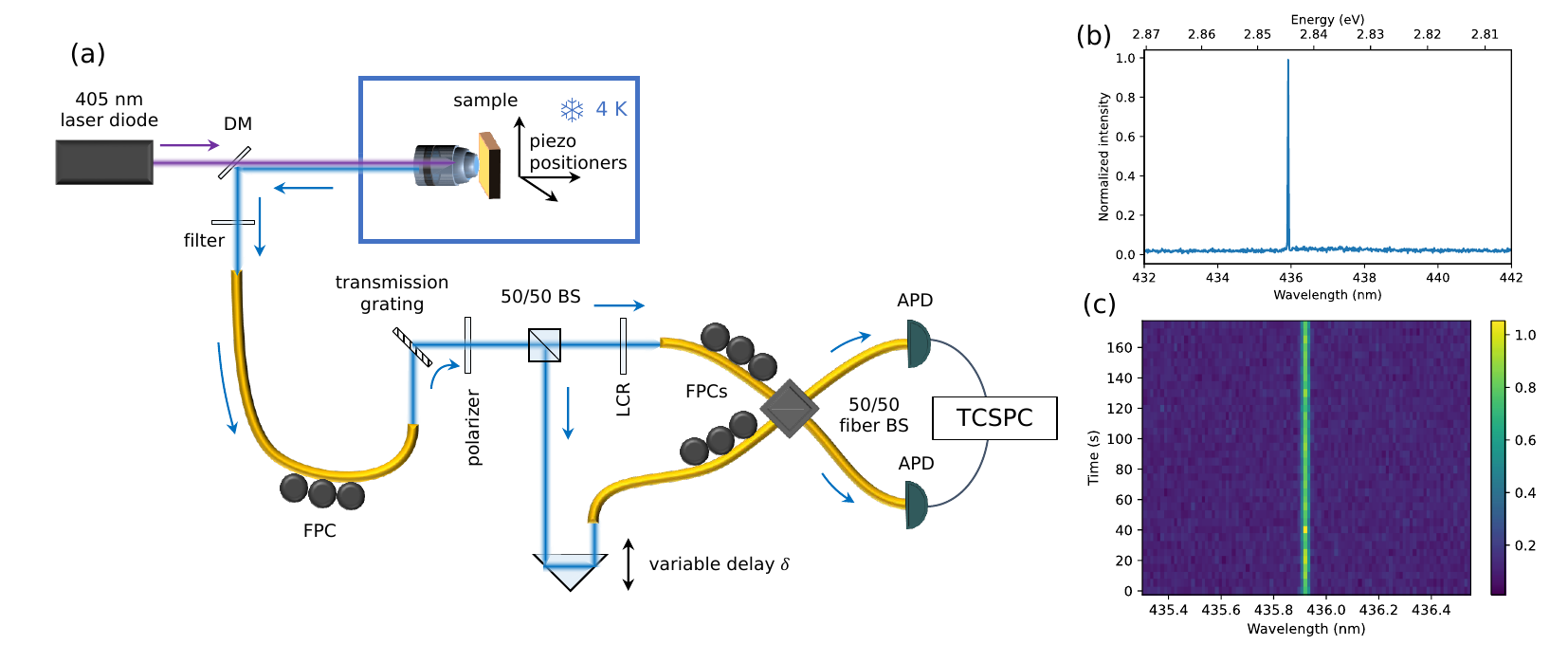}\\
  \caption{(a) Experimental setup. A cw laser diode excites the sample, which is inserted in a cryostat. The collected fluorescence light is channeled to a Mach-Zehnder interferometer with a variable arm. Light in the output arms is detected by APDs. DM: dichroic mirror; LCR: liquid crystal retarder; FPC: fiber polarization controler; TCSPC: time-correlated single photon counter. (b) PL spectrum measured using a grating spectrometer. (c) Time evolution of the PL emission during 180~s.}\label{setup}
\end{figure*}

Here, we perform PCFS of a B-center PL emission to characterize the homogeneous and SD-induced inhomogeneous broadening of the emission linewidth at timescales down to a microsecond. This allows us to establish the timescale and magnitude of the SD, as well as to quantify laser-induced dephasing in the non-resonant excitation regime. We show that the homogeneous linewidth (limited by pure dephasing) depends on the laser power and is close to the lifetime limit at powers $\leq 1$~mW. The linewidth is broadened at microsecond timescales by SD, whose amplitude depends only weakly on the laser power. The long-time linewidth is of the order of a gigahertz, comparable to that measured in resonant laser spectroscopy.

\section{II. Standard spectroscopy}
\label{II}

\subsection{A. Context and definitions}

The B center is modeled as a two-level system with ground state $|g\rangle$ and excited state $|e\rangle$ with a spontaneous emission rate $\Gamma_1 = 1/T_1$ and a pure dephasing rate $\Gamma_\phi = 1/T_\phi$. The decoherence rate is then $\Gamma_2 = 1 / T_2 = \Gamma_1/2 + \Gamma_\phi$, with $T_2$ refered to as the homogeneous coherence time of the emitter. A maximally coherent emitter with $\Gamma_\phi = 0$ has a lifetime-limited linewidth $2 \hbar\Gamma_1$ -- however, solid-state emitters often have finite pure dephasing rate $\Gamma_\phi$ due for instance to interactions with phonons. This leads to a degradation of the indistinguishability of the emitted photons, which goes as $T_2/2T_1$. Additionally, the emitter can exhibit a fluctuating emission energy $\hbar \omega_{ge}(t)$, often attributed to electrostatic coupling to nearby charge traps. In the general case, the dynamics of the emitter coherence $\rho_{eg} = \langle \sigma_{ge} \rangle$ can be expressed through the optical Bloch equations as:

\begin{equation}
\dot{\rho}_{ge} = i \omega_{ge}(t) \rho_{ge} - \Gamma_2 \rho_{ge}
\label{OBE}
\end{equation}

The field operators of the ZPL emission are proportional to the dipole operators of the emitter, \textit{i.e.} $a \propto \sigma_{ge}$~\cite{Loudon}. Therefore, the photon coherence is directly inherited from the emitter coherence. In particular, the first-order correlation function of the emitted light writes $g^{(1)}(\delta) = \left\langle \sigma_{eg}(t) \sigma_{ge}(t+\delta) \right \rangle$, and the power spectrum of the far-field emission in the steady state is given by the Fourier transform of $g^{(1)}(\delta)$. In the case where $\omega_{ge}$ is constant, the obtained homogeneous emission spectrum can be calculated from Eq.~\ref{OBE} and the quantum regression theorem~\cite{Loudon}, and is a Lorentzian of width $\hbar \Gamma_2$ centered around $\hbar \omega_{ge}$.

Upon random variations of $\hbar \omega_{ge}(t)$ --which we suppose to occur at timescales longer than $T_1$--, the time-integrated spectrum is then the average over all possible realizations of $\hbar \omega_{ge}(t)$. We note $S_\mathrm{inhom}^\infty(\omega)$ the probability distribution function of $\omega_{ge}(t)$. This inhomogeneous envelope is often Gaussian, although other spectral shapes are possible depending on the microscopic mechanisms at play, and can be for instance Lorentzian or multimodal. This randomization of the center frequency limits the number of indistinguishable photons that can be emitted. In the case of experiments that necessitate an integration time longer than the characteristic timescale of $\hbar \omega_{ge}(t)$, the randomization of the emitter transition energy is acting as an additional effective decoherence of characteristic timescale termed $T_2^*$. When $S_\mathrm{inhom}^\infty(\omega)$ is a Gaussian distribution of FWHM $\Delta \omega_\mathrm{inhom}$, $T_2^* = 4\pi\sqrt{\ln 2}/\Delta \omega_\mathrm{inhom}$.

The time-averaged inhomogeneous emission spectrum is then

\begin{align*}
S^\infty(\omega) &= \int d\omega' S_\mathrm{inhom}(\omega') \dfrac{\Gamma_2/\pi}{(\omega - \omega')^2 + \Gamma_2^2}\\
&= S_\mathrm{inhom}^\infty(\omega) * S_\mathrm{hom}(\omega)
\end{align*}
where we introduced the homogeneous lineshape $S_\mathrm{hom}(\omega)$ as a Lorentzian of width $\Gamma_2$ centered around zero frequency. The superscript $\infty$ indicates that integration is performed over a timescale exceeding all relevant dynamics.

When the variations of $\hbar \omega_{ge}(t)$ occur at too short timescales, standard spectroscopy techniques only allow access to $S^\infty(\omega)$ rather than $S_\mathrm{hom}(\omega)$ and $\hbar \omega(t)$, such that the identification of the homogeneous and inhomogeneous contributions can be challenging -- and information about the spectral dynamics is lost. In the following, we present the results of standard spectroscopy techniques in the case of a B center in hBN.

\subsection{B. Experimental setup and grating spectroscopy}

To allow insightful comparison between resonant and non-resonant excitation, we focus on an individual emitter that has already been extensively characterized under resonant excitation in a previous work~\cite{Gerard25}. Resonant laser spectroscopy provided a maximal coherence ($T_2 = 2 T_1$) at short times, and an inhomogeneous linewidth of 0.75~GHz due to SD occurring at timescales of milliseconds. The experimental setup is depicted on Fig.~\ref{setup}.

The sample is introduced in a closed-cycle cryostat, and excited by a continuous-wave (cw) 405~nm laser diode, which is focused on the sample by an objective of numerical aperture 0.8. The photoluminescence signal is collected by the same objective, and channeled to a single-mode fiber connected to a Mach-Zehnder interferometer. A transmission grating inserted in the input port ensures that only the zero-phonon-line (ZPL) photons are introduced in the interferometer. A polarizer fixes the input polarization before the first beamsplitter. A translation stage allows us to vary the time delay $\delta$ between the two arms from 0 to 1.3~ns. In the other arm, we place a liquid crystal retarder, which can be used to rotate the polarization by 90$^\circ$, allowing  us to switch from parallel to orthogonal polarization electronically to enable or disable any interference effects. The two arms recombine in a fiber beamsplitter, and avalanche photodiodes (APDs) detect the photons at the two output arms. This setup is suited for both FTS and PCFS. Alternatively, the collection fiber can be coupled to a grating spectrometer for performing standard dispersive optical spectroscopy. A photoluminescence spectrum of the emitter is provided on Fig.~\ref{setup}b and exhibits a narrow line at 435.92~nm. Its linewidth is limited by the spectrometer resolution of 150~$\mu$eV. The long-time evolution of the PL spectrum is shown on Fig.~\ref{setup}c, where no apparent fluctuations can be observed. This justifies further investigation using FTS and PCFS, presented in the following sections.

\subsection{C. Fourier-transform spectroscopy}
\label{subsection_FTS}

FTS allows to characterize the time-averaged spectrum $S^\infty(\omega)$ of the light emission with a much higher frequency resolution, only limited by the maximum optical delay that can be achieved. If the timescale of the inhomogeneous broadening mechanisms is longer than the emitter lifetime, the time-averaged emission spectrum $S^\infty(\omega)$ is then the convolution product of the homogeneous lineshape $S_\mathrm{hom}(\omega)$ and the time-averaged inhomogeneous envelope $S_\mathrm{inhom}^\infty(\omega)$, \textit{i.e.} $S^\infty = S_\mathrm{hom} * S_\mathrm{inhom}^\infty$. By measuring the interference fringes as a function of the optical time delay $\delta$, FTS accesses the first-order coherence function $g^{(1)}(\delta) = \widetilde{S^\infty}(\delta)$, where $\widetilde{\cdot} = \mathcal{F}[\cdot]$ stands for the Fourier transform.
 
\begin{figure}[t]
  \centering
  \includegraphics[width=0.8\linewidth]{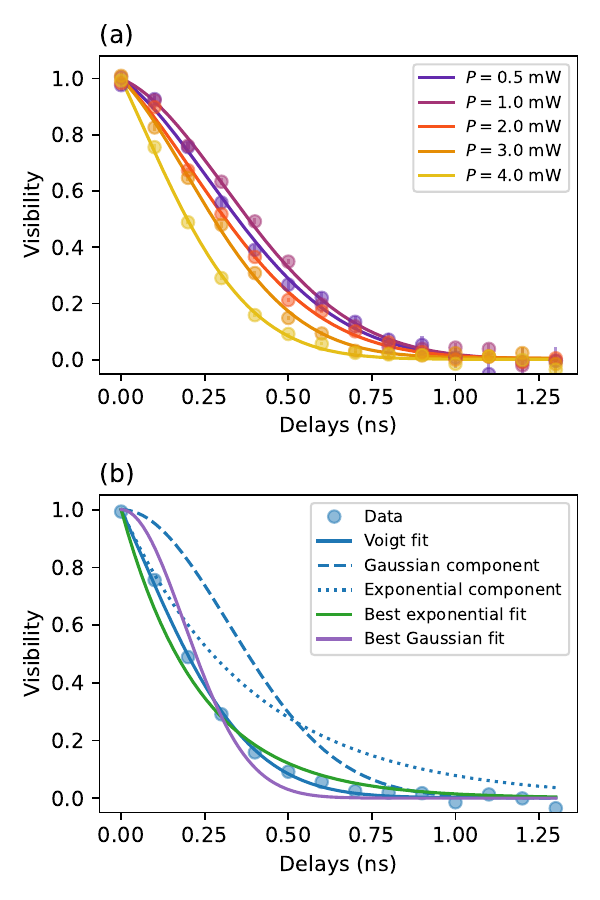}\\
  \caption{(a) Dots: Visibility as a function of the optical path length difference. Solid lines: Voigt fit to the data. (b) Dots: 4-mW data. Blue (resp. green, resp. purple) plain line: Voigt (resp. exponential, resp. Gaussian) fit to the data. Dashed (resp dotted) line: Exponential (resp. Gaussian) component of the Voigt decay profile.}\label{FTS}
\end{figure}

We perform FTS of the B center by recording interference fringes at varying delays $\delta$ from 0 to 1.3~ns. We measure the visibility $V(\delta) = \left|g^{(1)}(\delta)\right|$ at each position. The experimental procedure for extracting the fringe visibility is described in the Supplemental Material section S1~\cite{SM}. The result is plotted on Fig.~\ref{FTS}a for five different laser powers ranging from 0.3 to 4~mW (color dots). A clear decay of $V(\delta)$ can be observed at all powers. Additionally, the decay shows a distinct dependence on the laser power, with a faster decay at increasing power.

In the case of inhomogeneously broadened emitters, a common spectral shape is the Voigt profile~\cite{Armstrong67}, which corresponds to a Lorentzian homogeneous spectral shape $S_\mathrm{hom}(\omega) \propto 1/(\omega^2 + 1/T_2^2)$, with $T_2$ the emitter coherence time, fluctuating within a Gaussian envelope $S_\mathrm{inhom}^\infty(\omega) \propto \exp(-[T_2^*(\omega - \omega_0)/2]^2)$, where $T_2^*$ is the inhomogeneous coherence time and $\omega_0$ the center frequency of the distribution. This profile yields a visibility decay that is simply given by

\begin{equation}
V(\delta) = \exp\left(-\dfrac{\delta}{T_2}\right) \exp\left[
 - \left(
 \frac{\delta}{T_2^*}
 \right)^2
 \right]
 \label{Voigt_FTS}
\end{equation}

We fit the experimental visibility with Eq.~\ref{Voigt_FTS}. The plain lines on Fig.~\ref{FTS}a shows the fit results, which agree well with the experimental data at all powers. To verify that a simple exponential or Gaussian fit is insufficient to account for the decay shape, a comparison of different options of fitting functions is shown on Fig.~\ref{FTS}b for the 4~mW case, where it can be seen that the Voigt decay (blue line) fits the data (blue dots) significantly better than either a purely exponential fit (green curve) or a purely Gaussian fit (purple curve). Additionally, the two components (exponential in dotted line and Gaussian in dashed line) of the Voigt fit are plotted on the same graph. 

\begin{figure}[t]
  \centering
  \includegraphics[width=0.8\linewidth]{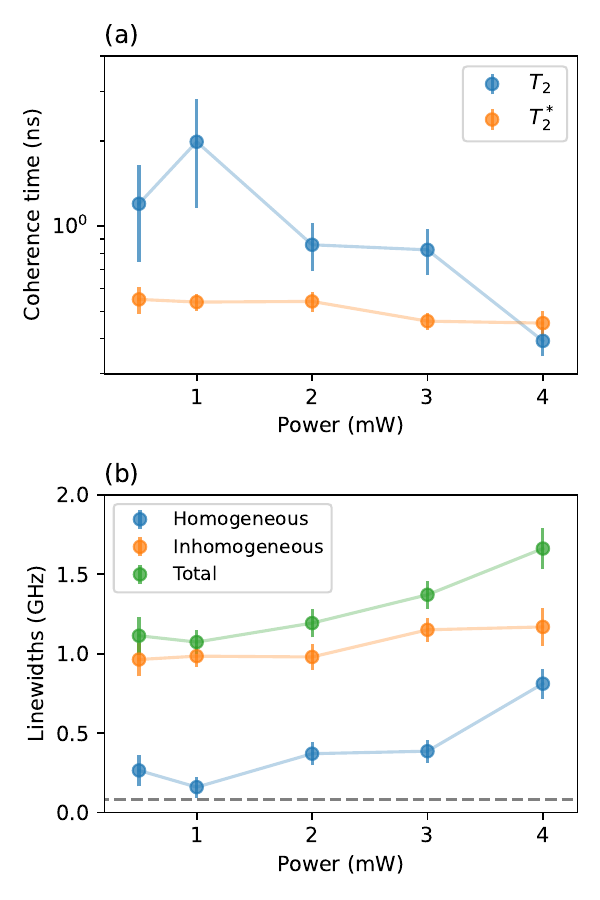}\\
  \caption{(a) Homogeneous ($T_2$, blue dots) and inhomogeneous ($T_2^*$, orange dots) coherence times as a function of the laser power. (b) Homogeneous (blue dots), inhomogeneous (orange dots) and total (green dots) linewidth as a function of the laser power. The dashed gray line indicates the Fourier limit.}\label{FTS_fit}
\end{figure}

Following Eq.~\ref{Voigt_FTS}, we attribute the exponential decay to the homogeneous contribution, with a coherence time $T_2$, and the Gaussian component to an additional effective decoherence with timescale $T_{2}^{*}$, corresponding to an inhomogeneous broadening of the spectrum at macroscopic times. Fig.~\ref{FTS_fit}a shows the fit parameters $T_2$ and $T_2^*$ as a function of the laser power. It can be observed that $T_2$ strongly depends on the power, decreasing from 2.0~ns at $P = 1$~mW down to 0.45~ns at $P = 4$~mW. This might originate from local heating induced by the excitation laser. At the lowest power (0.3~mW), $T_2$ also seems to be lower than the $P = 1$~mW case, although this apparent decrease might only be due to the large uncertainty. In turn, $T_2^*$ weakly depends on the laser power, with values between 0.45 and 0.55~ns at all powers. The fit parameters allow to directly infer the homogeneous linewidth $\Delta \omega_\mathrm{hom} = 2/T_2$ and the inhomogeneous linewidth $\Delta \omega_\mathrm{inhom} = 4 \sqrt{\ln 2}/T_2^*$, which are plotted on Fig.~\ref{FTS_fit}b, together with the total linewidth $\Delta \omega_\mathrm{tot}$ and the Fourier limit $1/T_1 = 2 \pi \times 82$~MHz inferred from a measurement of the PL decay time $T_1 = 1.9$~ns. The homogeneous linewidth is close to the Fourier limit at low powers (down to a factor~2) and increases as the power increases. The inhomogeneous contribution broadens the line up to above a gigahertz even at the lowest powers. To confirm that this Gaussian contribution originates from spectral diffusion, in the next step we perform PCFS, which gives access to the time dependence of this effective dephasing process. 

\section{III. Photon-correlation Fourier spectroscopy}
\label{III}

\subsection{A. Principle}

The principle and theory behind PCFS have been extensively described in prior work~\cite{Brokmann06, Coolen07}. Here, we briefly recall the main concepts and expose the results that are relevant for the following data analysis. We also introduce the concept of effective, time-dependent spectrum, which brings an intuitive general picture of the PCFS measurements. A detailed derivation of the relevant formulas and methods is provided in the Supplemental Material~\cite{SM}.

\subsubsection{1. PCFS contrast}

PCFS is based on the measurement of the second-order correlation function $g^{(2)}( \delta,\tau)$ from the output ports of an interferometer of variable optical delay difference~$\delta$. Here, $\tau$ denotes the delay time between photon detection events by the two APDs. The relative intensities in the two output ports 1 and 2 depend on the instantaneous emission wavelength and path length difference due to optical interference. In the case where there is constructive interference on port 1 -- and therefore destructive interference on port 2 --, the coincidence rate between photons that are shortly separated in time is reduced with respect to the average coincidence rate. Upon spectral fluctuations, the emission wavelength is randomized and the coincidence rate regresses towards the mean -- all the more so when the spectral fluctuations have a large amplitude or when the optical delay $\delta$ is long. In other words, PCFS measures the persistence time of constructive interference, which translates into antibunching in the cross-correlation signal. The dependence of the degree of antibunching on the optical delay $\delta$ provides information about the magnitude of the spectral jumps occurring during $\tau$.

The main advantage of PCFS is that the signal over noise ratio is not given by the average photon count rate during $\tau$ -- which is prohibitively low for small values of~$\tau$ -- but by the total integration time, which can be in principle arbitrary long. This provides a way to explore the fast dynamics of quantum emitters at timescales much shorter than the inverse count rate~\cite{Schimpf19}.

In a formal point of view, the coincidence rate can be expressed as~\cite{Brokmann06}:

\begin{equation}
g^{(2)}(\delta, \tau) = 1 - C_\phi(\tau) \mathcal{F}_\zeta \left[
p(\zeta, \tau)
\right]
\label{g2_eq}
\end{equation}
where $p(\zeta, \tau)$ is the spectral correlation function of the emitter (analyzed in the next section), and $C_\phi(\tau)$ is the time-delay-dependent contrast. It depends on the time variations of the interferometer phase difference $\phi(t)$ as $C_\phi(\tau) =  \frac{1}{2}\left\langle
\cos[ \phi(t + \tau) -\phi(t)] 
\right \rangle_t$, where $\langle  \cdots \rangle_t$ denotes the time average. $C_\phi$ matches the interferometric antibunching obtained with an ideal monochromatic source -- \textit{i.e.} $C_\phi(\tau) = 1 - g^{(2)}_\mathrm{class}(\tau)$, with $g^{(2)}_\mathrm{class}(\tau)$ the intensity autocorrelation function of the monochromatic source at the interferometer output~\cite{SM}.

A common approach to perform PCFS is to scan the interferometer optical delay at a constant velocity $2v$, \textit{i.e.} $\delta(t) = 2vt$, which yields $C_\phi(\tau) = 1/2 \cos(2\omega_0 v \tau / c)$~\cite{Brokmann06, Coolen07}. In the present work, we instead rely on free-running interference originating from fluctuating path length, such that detections separated by more than the fluctuation time (occurring at the timescale of a few 100s of milliseconds) are uncorrelated. In this case, $C_\phi(\tau)$ has a non-trivial dependence, and decreases from 0.5 at small $\tau$ to 0 at $\tau \gtrsim 1$~s, which sets the upper bound of the dynamics that can be observed with the setup. More details about the description of the free-running interference approach are provided in the Supplemental Material section~S1~\cite{SM}. 

The PCFS contrast of the emitter is then $C(\delta, \tau) = \mathcal{F}_\zeta \left[
p(\zeta, \tau) \right]$. For an emitter that does not exhibit intensity fluctuations at the relevant timescales, it can be obtained in both cases (constant velocity and free-running interference) by normalizing the degree of antibunching $1 - g^{(2)}(\delta, \tau)$ by its value at $\delta = 0$, \textit{i.e.}

\begin{equation}
C(\delta, \tau) = \mathcal{F}_\zeta \left[
p(\zeta, \tau) \right] = \dfrac{1 - g^{(2)}(\delta, \tau) }{1 - g^{(2)}(0, \tau) } 
\label{PCFS_contrast_eq}
\end{equation}
which removes the need for a specific characterization of $C_\phi(\tau)$. In the general case, the influence of any additional processes acting on the photon correlations (\textit{e.g.} intensity fluctuations, blinking or metastable states) can be corrected for using the orthogonal polarization signal~\cite{SM}. Note that for an ideal interferometer, the prefactor $C_\phi$ is independent of $\delta$, but the setup imperfections (alignment and collimation) can make the interference contrast decrease when $\delta$ increases, \textit{i.e.} $C_\phi^\mathrm{eff}(\delta, \tau) =C_\phi(\tau) V_\mathrm{class}^2(\delta)$, where $V_\mathrm{class}(\delta)$ stands for the optical-delay-dependent fringe visibility of a monochromatic source at delay $\delta$, and is accounted for by measuring the contrast decay of a single-mode laser (see Supplemental Material section S1.2~\cite{SM}).

\subsubsection{2. Spectral correlation and effective spectrum}

The spectral correlation function $p(\zeta, \tau)$ is the averaged cross-correlation of instantaneous spectra $s(\omega, t)$ separated by the delay $\tau$:

\[
p(\zeta, \tau) = \left\langle \int d\omega 
s(\omega, t) s(\omega + \zeta, t + \tau)
\right\rangle_t
\]



At short times, the quantity $p(\zeta, 0)$ matches the autocorrelation function of the homogeneous spectrum $S_\mathrm{hom}(\omega)$, and at long times, it identifies with the autocorrelation of the inhomogeneously broadened spectrum $S^\infty(\omega) = (S_\mathrm{hom} * S_\mathrm{inhom}^\infty)(\omega)$. This can be generalized at all delays $\tau$ by expressing $p(\zeta, \tau)$ as the autocorrelation function of an effective, time-dependent spectrum $S_\mathrm{eff}(\omega, \tau)$ (see Supplemental Material section S2~\cite{SM}). This spectral shape $S_\mathrm{eff}(\omega, \tau)$ consistently interpolates between the homogeneous, unbroadened lineshape at short times and the inhomogeneous, uncorrelated lineshape at long times, and contains the relevant information about the linewidth evolution associated with the SD dynamics.

In turn, this approach makes it possible to interpret the PCFS contrast as the autocorrelation function of this effective spectrum $S_\mathrm{eff}(\omega, \tau)$, which thus constitutes a convenient tool for the analysis and interpretation of the SD processes in PCFS. In particular, it allows for a proper definition of the time-dependent coherence time. In analogy to the time-averaged case, the effective spectrum $S_\mathrm{eff}(\omega, \tau)$ can be expressed as the convolution product of the homogeneous shape $S_\mathrm{hom}(\omega)$ and a time-dependent inhomogeneous envelope $S_\mathrm{inhom}(\omega, \tau)$. The latter carries information about SD and can be explicitly calculated from the SD kernel, which governs the spectral fluctuations~\cite{Reilly93, Delteil24, Gaignard25, SM}. More details about the definition and properties of $S_\mathrm{eff}(\omega, \tau)$ are provided in the Supplemental Material sections S2 and~S3~\cite{SM}.

The PCFS contrast $C(\delta, \tau)$ is obtained by Fourier transforming $p(\zeta, \tau)$ to the time domain (Eq.~\ref{PCFS_contrast_eq}), and can be written as

\begin{equation}
C(\delta, \tau) = \left|\tilde{S}_\mathrm{eff}(\delta, \tau) \right|^2  = \left|\tilde{S}_\mathrm{hom}(\delta)  \tilde{S}_\mathrm{inhom}(\delta, \tau) \right|^2
\label{PCFS_S_eff}
\end{equation} 

The PCFS contrast is the squared modulus of the Fourier transform of the effective spectrum $S_\mathrm{eff}$. This general result shows a strong analogy with the FTS approach measured in the steady state, where the visibility $V(\delta)$ is the modulus of the Fourier transform of the time-averaged spectrum

\begin{equation*}
V(\delta) = \left|\tilde{S}^\infty(\delta) \right|  = \left|\tilde{S}_\mathrm{hom}(\delta) \tilde{S}_\mathrm{inhom}^\infty(\delta) \right|
\end{equation*}

Eq.~\ref{PCFS_S_eff} allows for a direct derivation of the various particular cases treated in prior work~\cite{Coolen08, Coolen09, Prechtel13, Utzat19, Spokoyny20}.

\subsubsection{3. Example cases}

In the frame of microscopic SD models, $p(\zeta, \tau)$ and $S_\mathrm{eff}(\omega, \tau)$ can be expressed in terms of the SD kernel~\cite{Reilly93, Delteil24, Gaignard25}, so that $C(\delta, \tau)$ can be explicitly calculated. In this Section, we mention a few cases that are useful for the following data analysis. More details about the derivations are provided in the Supplemental Material Section~S3~\cite{SM}.

In the case of Gaussian broadening of a Lorentzian line, $\tilde{S}_\mathrm{hom}(\delta) = \exp(-\delta/T_2)$ and $\tilde{S}_\mathrm{inhom}(\delta,\tau) = \exp\left( - \delta^2/T_2^*(\tau)^2\right)$, with $T_2$ the coherence time and $T_2^*(\tau)$ the time-dependent effective inhomogeneous linewidth. $S_\mathrm{eff}(\omega, \tau)$ is therefore a Voigt profile. A well-known particular example of this case is the Ornstein-Uhlenbeck (OU) model, for which an analytical calculation from the OU SD kernel provides $T_2^*(\tau) = {T_2^*}^\infty / \sqrt{1 - \exp(-\tau/\tau_\mathrm{SD})}$. In general, the Gaussian broadening can exhibit more complex dynamics~\cite{Empedocles99,Plakhotnik10} such that $T_2^*(\tau)$ can deviate from this particular time dependence.
 
 The total PCFS contrast $C(\delta, \tau)$ is then the Fourier transform of a Voigt profile, and is given by

\begin{equation}
C(\delta, \tau) = \exp\left(-\dfrac{\delta}{T_2/2}\right) \exp\left[
 - \left(
 \frac{\delta}{T_2^*(\tau)/\sqrt{2}}
 \right)^2
 \right]
 \label{Voigt}
\end{equation}

Note that the exponential component decays with the timescale $T_2/2$ while the Gaussian component with $T_2^*/\sqrt{2}$, \textit{i.e.} the two relevant timescales are scaled by different factors with respect to the FTS decay counterpart of Eq.~\ref{Voigt_FTS}.

Another common case is that of a Lorentzian broadening~\cite{Coolen08, Coolen09, Schindler04, Kuhlmann13, Komza24}. In this case, the effective spectrum is a single Lorentzian of width $1/T_2 + 1/T_2^*(\tau)$, and the PCFS contrast decay at time $\tau$ is a single exponential of characteristic time $T_2/2 + T_2^*(\tau)/2$, consistently with exact calculations from a microscopic model~\cite{Coolen07,Coolen08}. 

Finally, in the case of discrete Poissonian random jumps within a Gaussian envelope~\cite{Utzat19, Spokoyny20, Gerard25, Wolters13}, the PCFS decay is the sum of a Lorentzian and a Voigt decay, both having fixed timescales but varying relative weights: 

\begin{align}
C(\delta, \tau)  &= a(\tau) \exp\left(-\dfrac{\delta}{T_2/2}\right) \nonumber
 \\
&+ (1-a(\tau)) 
\exp\left(-\dfrac{\delta}{T_2/2}\right)
\exp\left[
 - \left(
 \frac{\delta}{T_2^*/\sqrt{2}}
 \right)^2
 \right]
 \label{GRJ}
\end{align}

\subsection{B. Experimental results}

\begin{figure}[t]
  \centering
  \includegraphics[width=0.9\linewidth]{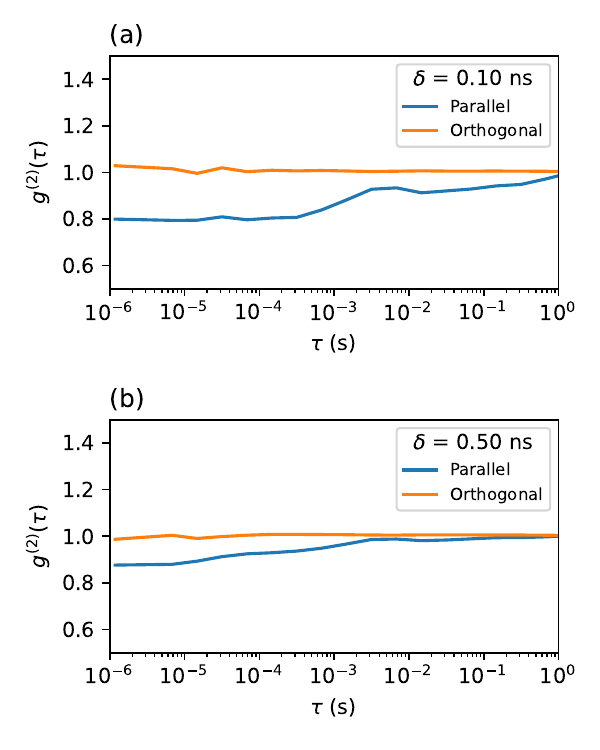}\\
  \caption{Interferometric $g^{(2)}(\delta, \tau)$ measured at two different optical delays, $\delta = 0.1$~ns (a) and $\delta = 0.5$~ns. Blue (orange) curves: Parallel (orthogonal) polarization. }\label{g2}
\end{figure}

The experimental PCFS contrast is obtained by measuring the $g^{(2)}(\delta, \tau)$ of the PL emission for each position of the delay stage $\delta$ from 0 to 1.3~ns, with the polarization being fixed to be either parallel (yielding $g^{(2)}_\parallel(\delta, \tau)$) or orthogonal ($g^{(2)}_\perp(\delta, \tau)$). Fig.~\ref{g2}a and b show two examples of $g^{(2)}(\delta, \tau)$ measured at two given optical delays $\delta = 0.1$~ns (Fig.~\ref{g2}a) and $\delta = 0.5$~ns (Fig.~\ref{g2}b). The binning for the time delay $\tau$ is logarithmic, with 3~bins per decade, except at the shortest delays ($\tau \leq 10$~$\mu$s) where the bins are made wider to increase the signal over noise ratio. In the orthogonal case, where any interference effects are suppressed, we obtain $g^{(2)}_\perp(\delta, \tau) = 1$ at all delays, thereby testifying of the absence of blinking. On the other hand, $g^{(2)}_\parallel(\delta, \tau)$ displays interferometric antibunching~\cite{Lebreton13}, with an amplitude that decreases with increasing $\tau$. Its non-trivial dependence on $\tau$ is related to the complex behavior of $C_\phi(\tau)$ (see Supplemental Material section~S1.3). At $\delta = 0.5$~ns, the antibunching is reduced with respect to the short delay case, in particular at large values of $\tau$, foreshadowing the complex dependence of the coherence on $\tau$ and $\delta$. 

\begin{figure*}[t]
  \centering
  \includegraphics[width=\linewidth]{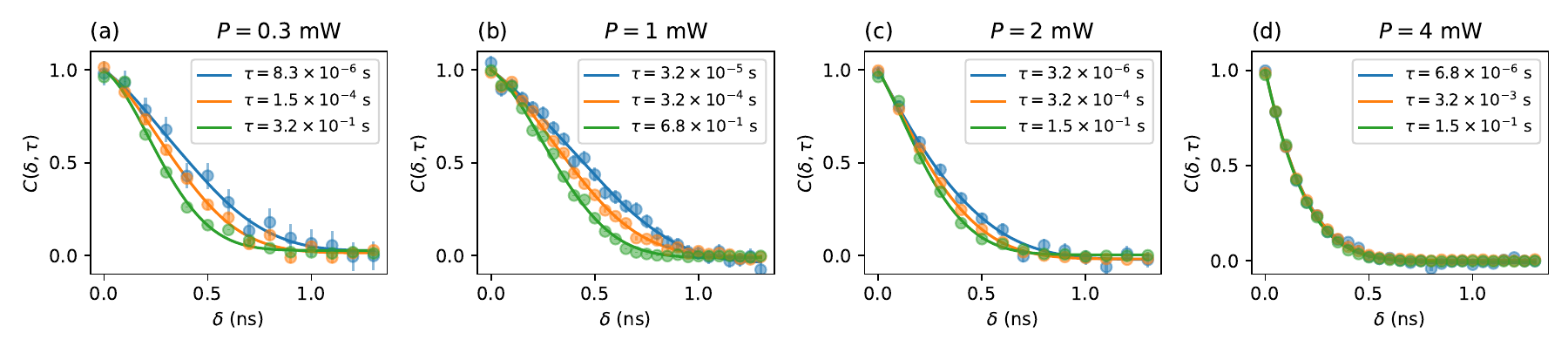}\\
  \caption{Color dots: PCFS contrast $C(\delta, \tau)$ for three different values of $\tau$ measured at different laser powers: 0.3~mW (a), 1~mW (b), 2~mW (c) and 4~mW (d). The solid lines are fits to the data using Eq.~\ref{Voigt}.   }\label{PCFS_contrast}
\end{figure*}

The measurement is repeated for four values of the excitation laser power. The PCFS contrast $C(\delta, \tau)$ is extracted from this data as $C(\delta, \tau) =\frac{1 - g^{(2)}_\parallel(\delta, \tau)}{1 - g^{(2)}_\parallel(0,\tau)}$. 
Fig.~\ref{PCFS_contrast} (dots) shows the contrast $C(\delta, \tau)$ as a function of $\delta$ for three values of $\tau$ chosen to reveal the variation of the decay shape and timescale when $\tau$ varies. At this stage, several insightful observations can be drawn. At the lowest powers (0.3~mW and 1~mW), the decay shapes have a visible Gaussian character, and the decay timescale decreases as $\tau$ increases. This is attributed to a shortening of the time-dependent coherence time $T_2^*(\tau)$ induced by fast spectral diffusion, which randomizes the emission wavelength. At higher power, the lineshapes are more exponential and exhibit a weaker dependence on the time delay $\tau$, consistently with a predominance of pure dephasing with increasing power, as hypothesized in Section~II.

\subsection{C. Data analysis}
\label{C}

\subsection{1. Continuous diffusion model}

\begin{figure}[t]
  \centering
  \includegraphics[width=0.8\linewidth]{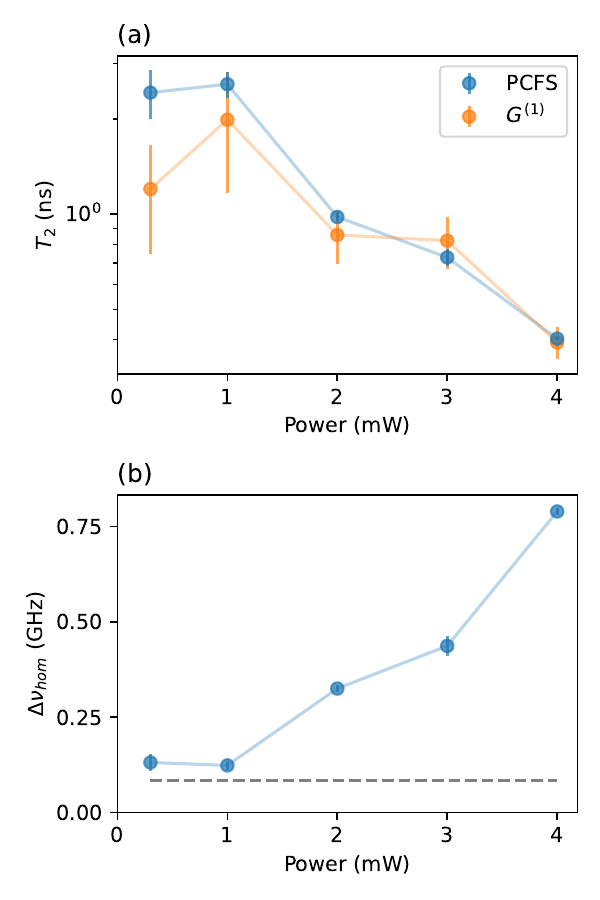}\\
  \caption{(a) Coherence time $T_2$ extracted from the decay of the PCFS contrast (blue dots) and from the FTS decay (orange dots) as a function of the laser power. (b) Homogeneous linewidth extracted from the PCFS contrast as a function of the laser power.}\label{PCFS_expo}
\end{figure}

In this section, we make the assumption of a continuous Gaussian broadening, where the effective inhomogeneous spectrum is a Gaussian function with a time-dependent linewidth, associated with the inhomogeneous coherence time $T_2^*(\tau)$. At this stage, we do not make any hypothesis on the time dependence of $T_2^*(\tau)$. We therefore fit the PCFS contrast $C(\delta, \tau)$ with the expression given in Eq.~\ref{Voigt}. The homogeneous $T_2$ is a shared fit parameter at each power, while $T_2^*(\tau)$ is a $\tau$-dependent fit parameter, consistently with the model. The fit results are shown as plain lines on Fig.~\ref{PCFS_contrast}, where a good agreement with the data can be observed at all powers and delays. We first focus on the exponential component. The associated coherence time $T_2$ is plotted on Fig.~\ref{PCFS_expo}a (blue dots), together with the results from FTS obtained in Section~II.B for comparison. The two approaches provide consistent values for $T_2$, with the PCFS approach yielding smaller uncertainties. The associated homogeneous linewidth is plotted on Fig.~\ref{PCFS_expo}b. At 0.3~mW (1~mW), we obtain $T_2 = 0.13 \pm 0.02$~GHz ($T_2 = 0.12 \pm 0.01$~GHz), which is only 60~\% above the Fourier limit.

\begin{figure}[t]
  \centering
  \includegraphics[width=0.8\linewidth]{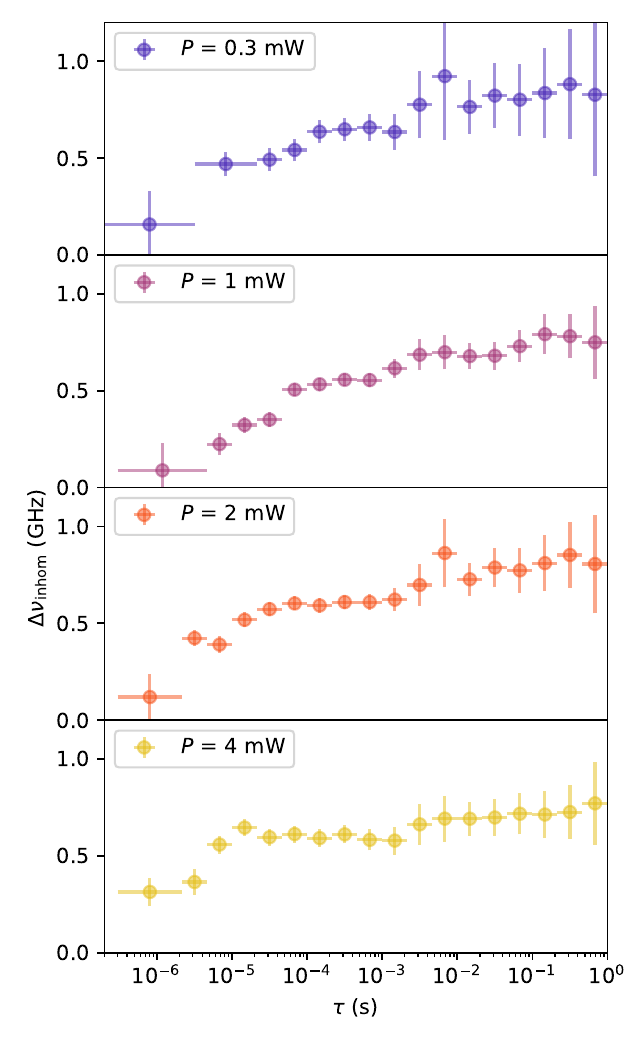}\\
  \caption{Time-dependent inhomogeneous broadening as a function of $\tau$ at four different laser powers.}\label{PCFS_gauss}
\end{figure}

We then consider the time-dependent inhomogeneous linewidth $\Delta \omega_\mathrm{inhom} = 4 \sqrt{2}/ T_2^*(\tau)$, deduced from the Gaussian contribution to the fit. Fig.~\ref{PCFS_gauss} shows the fit results as a function of $\tau$ at four different powers. All exhibit a common trend, with $\Delta \omega_\mathrm{inhom}(\tau)$ having a low value at short $\tau$ (\textit{i.e.} $\tau \sim$~1~$\mu$s) and monotonically increasing at longer delays. The increase is sharper at about $\tau \sim$~10~$\mu$s, and then only presents a slight incline.

\begin{figure}[t]
  \centering
  \includegraphics[width=0.8\linewidth]{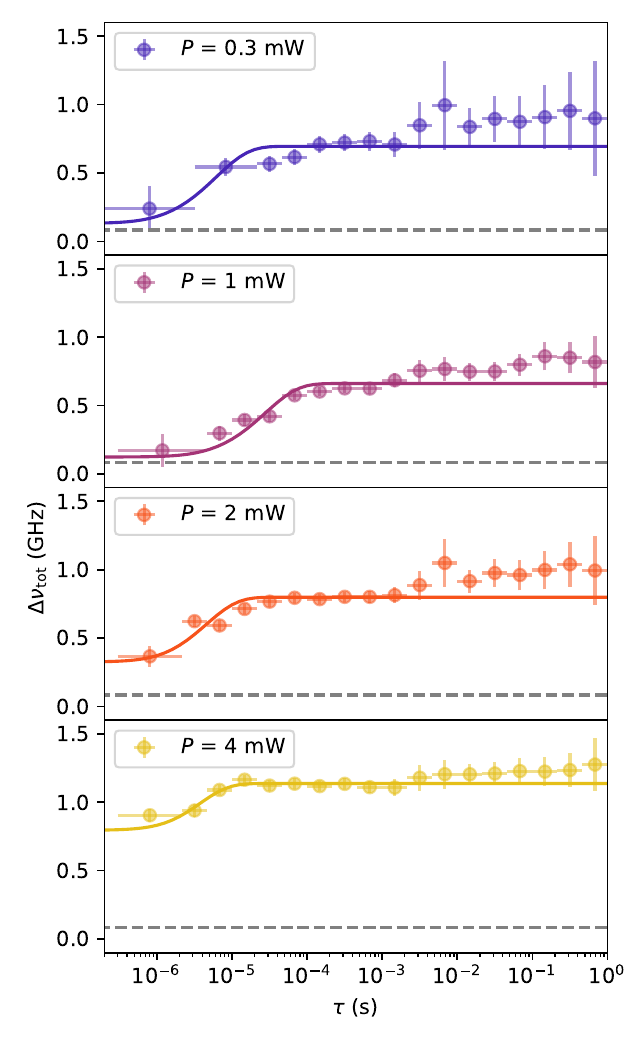}\\
  \caption{Dots: Time-dependent linewidth as a function of $\tau$ at four different laser powers. Plain lines: Fit to the data using the OU model.}\label{PCFS_totals}
\end{figure}

Accounting for both the homogeneous and the inhomogeneous broadening contributions, we plot on Fig.~\ref{PCFS_totals} the total linewidth as a function of $\tau$ for the four same values of the laser power (color dots). The gray dashed line depicts the Fourier limit of 82~MHz. At low power, the total linewidth is close to this lower bound for $\tau < 10$~$\mu$s, with $\Delta \omega_{tot}/2\pi = 0.24 \pm 0.16$~GHz at 0.3~mW, and $\Delta \omega_{tot}/2\pi = 0.17 \pm 0.12$~GHz at 1~mW. The linewidth broadens when $\tau $ increases, as per the behavior of the inhomogeneous component. As mentioned in Section~III.A.3, a particular case of Gaussian broadening is the OU model, where the linewidth evolves as $\Delta \omega_\mathrm{inhom} = \Delta \omega_\mathrm{inhom} ^\infty \sqrt{1 - \exp(-\tau/\tau_\mathrm{SD})}$, with $\tau_\mathrm{SD}$ is the SD correlation time. The plain lines show a fit with the OU linewidth broadening, with $\tau_\mathrm{SD}$ is a fit parameter. As can be observed on Fig.~\ref{PCFS_totals}, the rapid increase of the linewidth at about $10^{-5} - 10^{-4}$~$\mu$s is well captured, but the slow increase at times of milliseconds and longer would require to invoke an additional mechanism governing the long-time diffusion.

\subsection{2. Discrete spectral jumps}

\begin{figure}[t]
  \centering
  \includegraphics[width=0.8\linewidth]{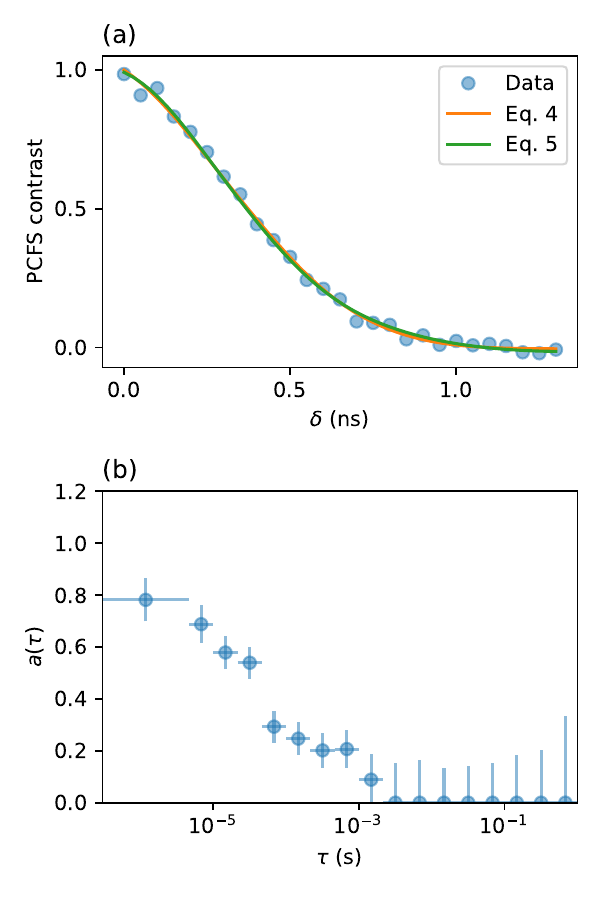}\\
  \caption{(a) Blue dots: PCFS contrast as a function of the delay $\delta$ at 1~mW. Orange curve: Fit using Eq.~\ref{Voigt}. Green curve: Fit using Eq.~\ref{GRJ}. (b) Relative weight of the homogeneous component as a function of time delay $\tau$.} \label{compar_OU_GRJ}
\end{figure}

There has been some investigations on the various signatures of the continuous vs.\ discrete character of the SD mechanism of solid-state emitters~\cite{Beyler13, Utzat19, Spokoyny20, Gerard25, Wolters13, Delteil24}. In PCFS, the two classes of SD yield different PCFS decays~\cite{Beyler13, Spokoyny20}, respectively associated with Eqs.~\ref{Voigt} and~\ref{GRJ}. We emphasize that the two models provide the exact same lineshapes at both short and long timescales, and differ only by the way the effective spectrum $S(\omega,\tau)$ transitions from the homogeneous to the inhomogeneous lineshape. In our case, we observe that a continuous approach is in excellent agreement with the whole dataset. However, given the decay times associated with the Gaussian and Lorentzian components of the PCFS contrast are not very far apart, the decay shapes can actually be well fitted by both Eqs.~\ref{Voigt} and~\ref{GRJ}. This can be seen on Fig.~\ref{compar_OU_GRJ}a with a representative example of a PCFS contrast decay. The whole dataset at $P = 1$~mW is fitted with Eq.~\ref{GRJ}, with $T_2$ and $T_2^*$ being shared fit parameters. The extracted time-dependent relative weight $a(\tau)$ of the homogeneous component is in line with expectations, with a close to homogeneous lineshape at short times, and a monotonic transition to the inhomogeneously broadened spectrum line when $\tau$ increases. We also observe an inflexion point at $\sim 3 \times 10^{-5}$~s, matching that of $\Delta \omega_{tot}(\tau)$ in the continuous diffusion approach. Therefore, a clear discrimination between the two classes of mechanisms would require to identify additional signatures~\cite{Delteil24,Gerard25}. Yet both approaches yield consistent results in terms of homogeneous and inhomogeneous contributions, time-dependent linewidth and lineshape, as well as SD timescales.

\section{IV. Conclusion}

We presented an exhaustive investigation of the photoluminescence coherence properties of an individual B center in hBN using PCFS. We identified two contributions to the linewidth, a power-dependent homogeneous broadening, associated with thermal dephasing, and a time-dependent inhomogeneous linewidth broadening, originating from SD at timescales of about 10~$\mu$s. At low power and short time, the linewidth is a factor 2 to 3 away from the Fourier limit. This is already sufficient to obtain a sizable degree of indistinguishability of consecutive photons using post-selection~\cite{Fournier23PRA} or Purcell enhancement~\cite{Gerard24}. Additionally, this level of coherence could readily enable applications to quantum sensing based on single-photon interferometry~\cite{Staunstrup24}, where the use of non-resonant excitation could be an asset for a flexible approach. A high indistinguishability at all time delays could be obtained by inserting the emitter in a cavity. A Purcell factor of 20 would be sufficient for obtaining a Hong-Ou-Mandel visibility of 90~\% when interfering photoluminescence photons separated by macroscopic time delays -- or equivalently, from remote emitters of identical center frequencies, undergoing independent spectral fluctuations. The PCFS approach employed here, which does not require interferometric stabilization, is readily applicable to a broad range of solid-state emitters. We anticipate that it will facilitate the identification and mitigation of decoherence mechanisms, thereby advancing the performance of quantum emitters for practical applications. 

\section{Acknowledgments}
The authors thank D. G\'erard and J. Claudon for fruitful discussions. This work is supported by the French Agence Nationale de la Recherche (ANR) under references ANR-21-CE47-0004.

\section{Data availability}
The data that support the findings of this article are openly available~\cite{zenodo}.

\pagebreak
~
\newpage

\onecolumngrid
\begin{center}
  \textbf{\large Supplemental Material\\~\\Photon correlation Fourier spectroscopy of a B center in hBN}\\[.2cm]
  Aymeric Delteil, St\'ephanie Buil, Jean-Pierre Hermier\\[.1cm]
  {\itshape \small Universit\'e Paris-Saclay, UVSQ, CNRS,  GEMaC, 78000, Versailles, France. \\
{\color{white}--------------------} aymeric.delteil@usvq.fr{\color{white}--------------------} \\}

\end{center}

\setcounter{equation}{0}
\setcounter{figure}{0}
\setcounter{table}{0}
\setcounter{page}{1}
\renewcommand{\theequation}{S\arabic{equation}}
\renewcommand{\thefigure}{S\arabic{figure}}

\section{S1. $g^{(1)}$ and interferometric $g^{(2)}$ under free-running interference}

\subsection{S1.1 Random fluctuations of the Mach-Zehnder interferometer}

\begin{figure}[h]
  \centering
  \includegraphics[width=1.0\linewidth]{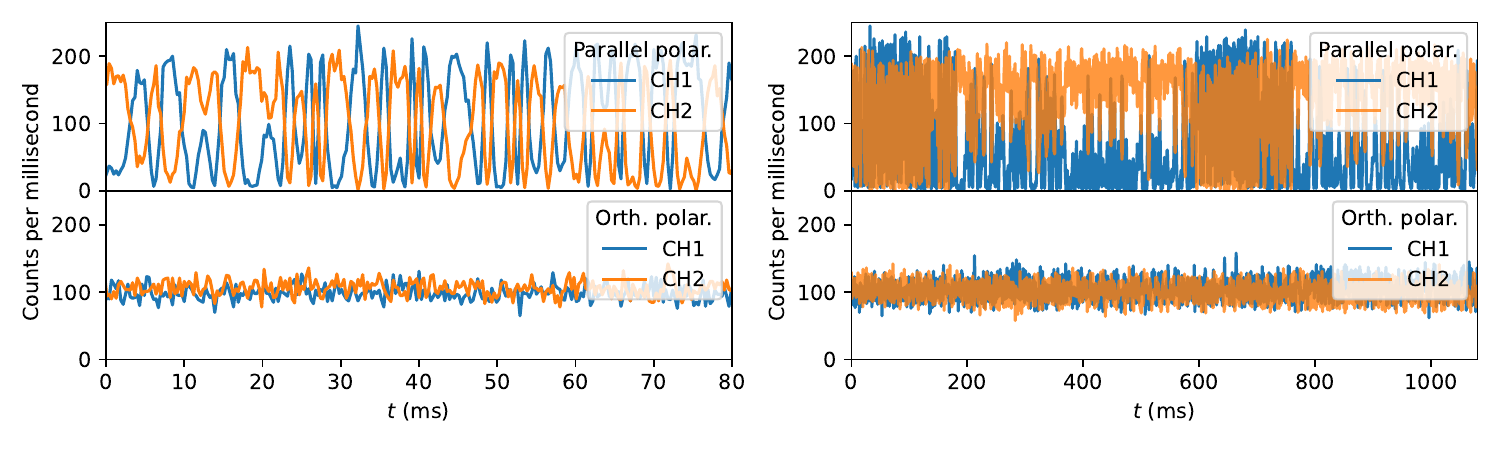}\\
  \caption{Signal measured at the output of the interferometer in APD1 (blue curve) and APD2 (orange curve) in the parallel (top) and orthogonal (bottom) configuration. Each column corresponds to a different time range.}\label{timetraces}
\end{figure}

Fig.~\ref{timetraces} show the time trace of a monochromatic laser as measured by the two APDs at the output of the interferometer, respectively $I_1(t)$ and $I_2(t)$ for parallel (top panel) and orthogonal (bottom panel) polarizations. In the parallel configuration, anticorrelated fluctuations can be observed, at timescales of a few milliseconds to a few hundreds of milliseconds. They originate from single-photon interference in the presence of fluctuations of the path length difference. In the orthogonal configuration, all interference effects are suppressed and the signals are constant (bottom panel).

\begin{figure}[h]
  \centering
  \includegraphics[width=0.6\linewidth]{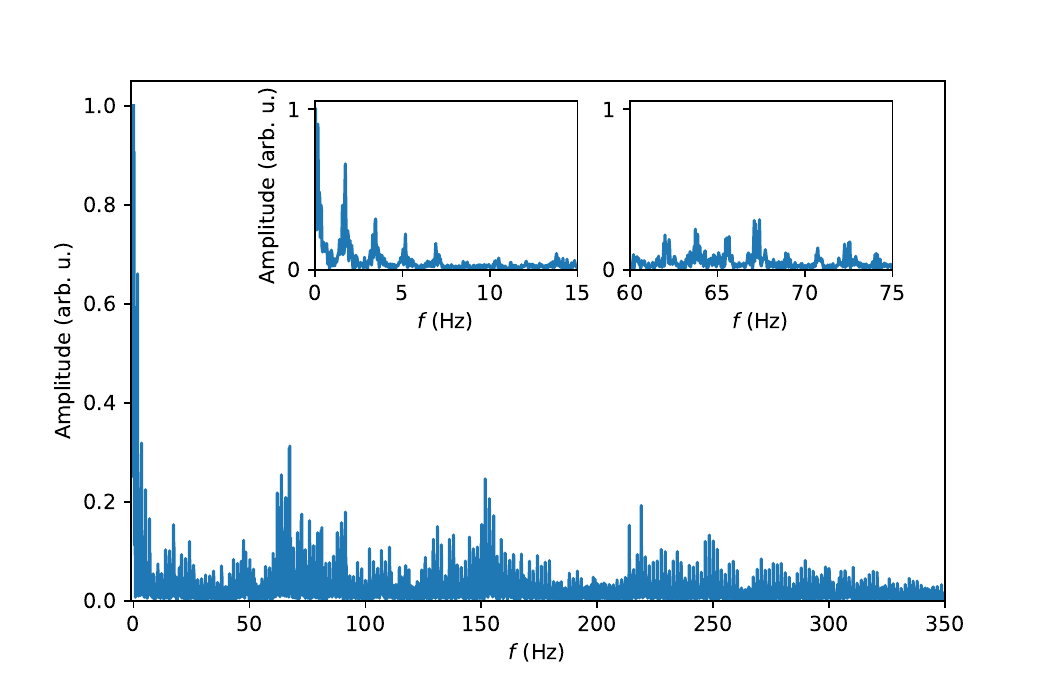}\\
  \caption{Fourier transform of the interferometric signal in APD1. Insets: zoom-in on specific frequency ranges.}\label{FTtimetraces}
\end{figure}

The fluctuations of the path length difference arise from mechanical vibrations and drift of the optical components, including slow thermal fluctuations of the fiber optical length and faster vibrations transmitted from the rotary valve of the nearby closed-cycle cryostat. More insight can be gained by performing the Fourier transform of the time traces of Fig.~\ref{timetraces}. The result is shown on Fig.~\ref{FTtimetraces}. Beyond the low-frequency peaks, sizable weight is located around 64~Hz and 155~Hz. All the finite-frequency components are distributed around harmonics of the rotary valve period of 1.7~Hz.

\begin{figure}[h]
  \centering
  \includegraphics[width=0.8\linewidth]{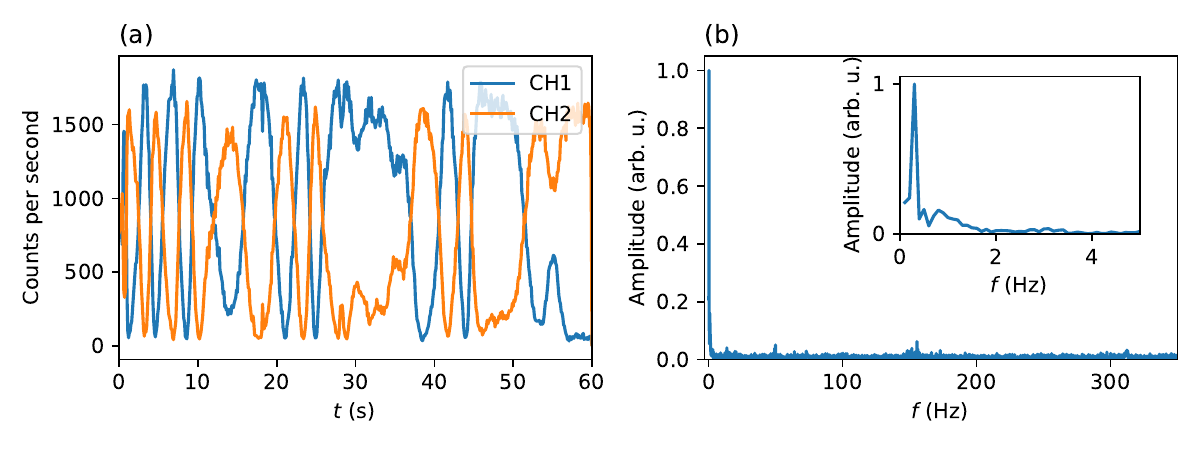}\\
  \caption{(a) Free-running interference of the laser in the parallel polarization configuration when the closed-cycle cryostat is off. (b) Fourier transform of the signal, which exhibit a single peak at about 0.3~Hz.}\label{Cryo_off}
\end{figure}

Switching off the cryostat, the interferometric signal is left with only the slow-fluctuating component at about 0.3~Hz, originating from mechanical and thermal variations in the path length difference. Fig.~\ref{Cryo_off} shows a timetrace and its Fourier transform under such conditions. Although the fluctuation timescale is much slower than in the case where the rotary valve is on, the presented experiments would also have been doable in those conditions. Indeed, the condition to be met is that the interferometer fluctuation timescale is longer than the timescales of the physical processes under study, but shorter than the total integration time.

\newpage

\subsection{S1.2 First-order correlation measurements}

To measure the first-order correlation function $g^{(1)}(\delta)$, the visibility has to be inferred at each delay~$\delta$. However, interference fringes under free running path length fluctuations, such as those shown on Fig.~\ref{timetraces}, cannot be directly fitted by a sine function. Instead, the interference visibility is derived from the statistics of the timetrace.

\begin{figure}[h]
  \centering
  \includegraphics[width=0.9\linewidth]{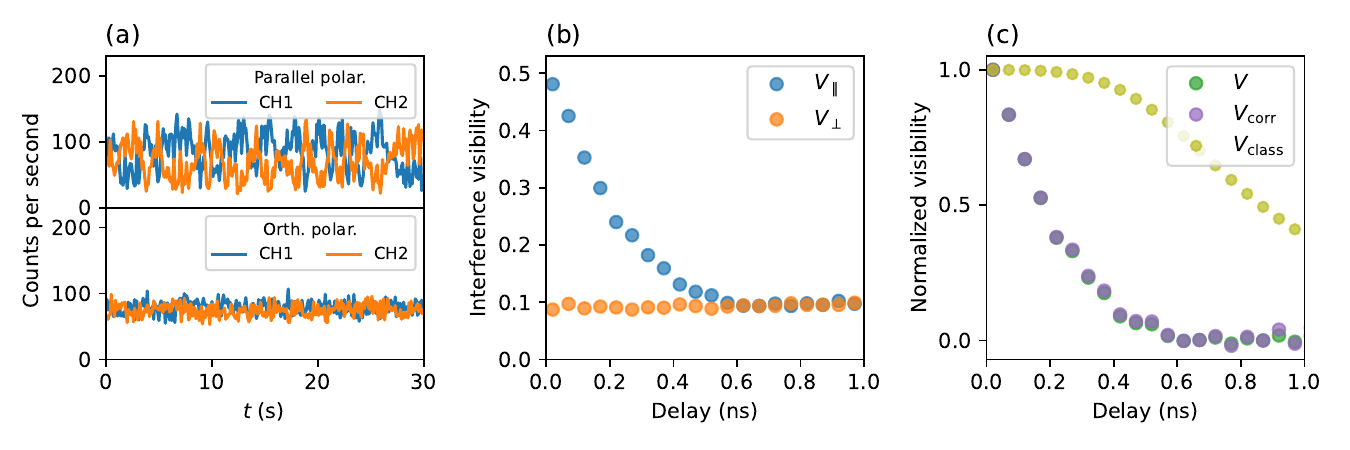}\\
  \caption{(a) Timetraces of the PL signal at the output of the interferometer in the parallel (top) and orthogonal (bottom) polarization configuration. (b) Visibilities $V_\parallel(\delta)$ (blue dots) and $V_\perp(\delta)$ (orange dots) extracted from the emitter PL in both polarization configurations. (c) Green dots: resulting raw visibility $V(\delta)$. Purple dots: visibility after correction by the maximal laser visibility. Yellow dots: laser visibility $V_\mathrm{class}$. The visibilities are normalized by their value at $\delta = 0$.}\label{g1}
\end{figure}

Noting $I_1(t)$ (resp. $I_2(t)$) the signal measured in APD1 (resp. APD2), the visibility can be expressed as 
\begin{equation}
V = \sqrt{2}\ \mathrm{std}\left(\dfrac{I_1(t) - I_2(t)}{I_1(t) + I_2(t)} \right) 
\label{vis_formula}
\end{equation}
where std stands for standard deviation. This method, valid for path length fluctuations larger than the wavelength, is robust to fluctuations of the input signal intensity. However, in the case of the photoluminescence signal, there are other sources of fluctuations of the output signal -- in particular shot noise, which is relevant at low count rates. Therefore, to isolate the contribution of interference, the same process is performed in both the parallel and the orthogonal polarization configuration. An example of such measurement is provided on Fig.~\ref{g1}. PL timetraces are measured at each delay in both polarizations. An example of such timetrace is shown on Fig.~\ref{g1}a. As for the laser case of Fig.~\ref{timetraces}, fluctuations at various timescales can be observed in the parallel configuration, originating from free-running interference. On the other hand, in the orthogonal configuration, the signals are constant. The associated visibilities are calculated based on Eq.~\ref{vis_formula}, and are shown Fig.~\ref{g1}b. The contribution attributed purely to interference is $V(\delta) = V_\parallel(\delta)-V_\perp(\delta)$, and is shown as green dots on Fig.~\ref{g1}c. Additionally, $V(\delta)$ is normalized by the visibility of the single-mode laser (of linewidth 100~kHz, thus with a coherence length much larger than the size of the interferometer) $V_\mathrm{class}(\delta)$ measured in the same conditions (yellow dots on Fig.~\ref{g1}c). $V_\mathrm{class}(\delta)$ sets the reference of the maximum achievable interference visibility at delay $\delta$, limited by mode overlap and polarization matching, and decreases when $\delta$ increases due to imperfect alignment and collimation. The result of this correction is $V_\mathrm{corr}(\delta) = V(\delta)/V_\mathrm{class}(\delta)$, and is shown as purple dots on Fig.~\ref{g1}c. Note that $V_\mathrm{class}(\delta)$ does not vary much in the range where the emitter coherence is sizable ($\delta \leq 0.5$~ns), and therefore the corrected visibility $V_\mathrm{corr}(\delta)$ differs only slightly from $V(\delta)$, as can be seen on Fig.~\ref{g1}c. The corrected visibility $V_\mathrm{corr}(\delta)$ is simply termed $V(\delta)$ in the main text for simplicity.

\subsection{S1.3 Second-order correlation measurements}

\subsubsection{S1.3.1 Monochromatic source}

In the case of a monochromatic input signal of constant intensity $I_0$, the two signals $I_1(t)$ and $I_2(t)$ measured at the output ports of the interferometer can be expressed as $I_1(t) = I_0 [1 + \cos\phi(t)]/2$ and $I_2(t) = I_0 [1 - \cos\phi(t)]/2$, where $\phi(t)$ is the fluctuating phase difference between the two arms.

The second-order cross-correlation functions at the output of the interferometer, fixed at a given path length difference $\delta$, are 

\begin{equation}
g^{(2)}_{12}(\tau) = 
\dfrac{
\left\langle
I_1(t) I_2(t + \tau)
\right\rangle_t
}{
\left\langle
I_1(t) \right\rangle_t
\left\langle
I_2(t)
\right\rangle_t
}
= 1 - C_\phi(\tau)
\label{crosscorr}
\end{equation}
\[
g^{(2)}_{21}(\tau) = g^{(2)}_{12}(\tau)
\]
where $C_\phi(\tau) = \frac{1}{2}\left\langle
\cos[ \phi(t + \tau) -\phi(t)] 
\right \rangle_t$. The degree of antibunching $C_\phi(\tau)$ takes values between $-1/2$ and $1/2$, and varies from $1/2$ at $\tau = 0$ to 0 at $\tau$ larger than the timescale of the phase fluctuations. Note that $g^{(2)}_{12}(\tau) \geq 0.5$ for the interferometric cross-correlation function of a classical light source~\cite{Lebreton13}.

Instead of using free-running fluctuations, some prior experiments rely on a small deterministic variation of the path length $\delta(t) = \delta_0 + 2 v t$ around each delay $\delta_0$, such that  $\phi(t) = 2 \omega_0 v t/c $~\cite{Brokmann06, Coolen07}. In this case, we obtain  $ C_\phi(\tau) = \frac{1}{2}\cos 2 \omega_0 v \tau/c$.

Note that in both cases $\delta$ is absent from the expression of $g^{(2)}_{12}$, since the source is supposed perfectly monochromatic (no contrast decay with increasing path length difference), and the time-varying phase runs over much more than a full period -- therefore the mean path length difference is just a phase offset that is cancelled out in $C_\phi$ when averaging over the phase variations.

Measuring the second-order correlation at the output of a single output port provides the intensity autocorrelation of the interference fringes
 \[
g^{(2)}_{11}(\tau) = 
\dfrac{
\left\langle
I_1(t) I_1(t + \tau)
\right\rangle_t
}{
\left\langle
I_1(t) \right\rangle_t^2
}
= 1 + C_\phi(\tau)
\]
 \[g^{(2)}_{22}(\tau) = g^{(2)}_{11}(\tau)\]

\begin{figure}[h]
  \centering
  \includegraphics[width=0.8\linewidth]{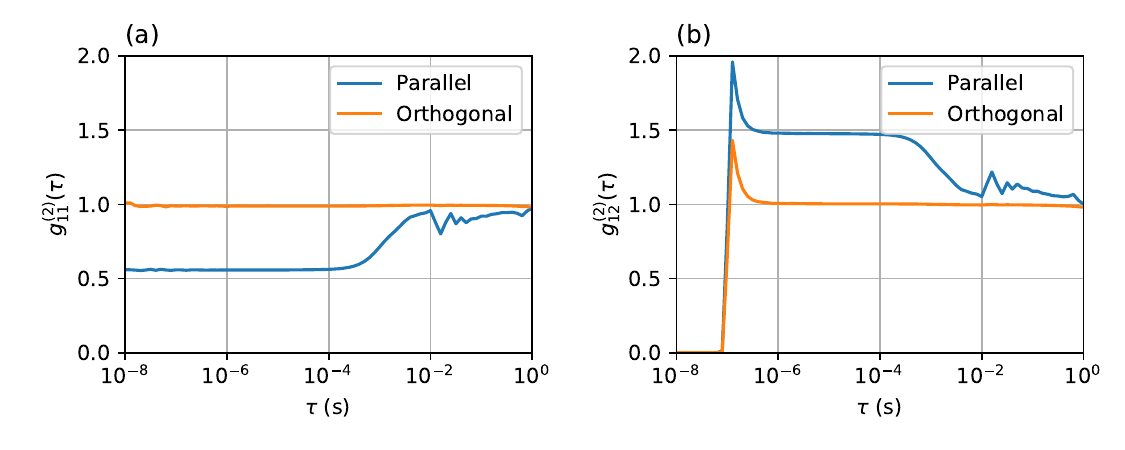}\\
  \caption{(a) Interferometric cross-correlation and (b) autocorrelation of the single-mode laser in parallel (blue) and orthogonal (orange) polarization configurations.  
  }
  \label{g2_laser}
\end{figure}

Fig.~\ref{g2_laser}a displays the experimental interferometric cross-correlation of our single-mode laser for both polarizations, at $\delta = 0$. While for orthogonal polarization no antibunching is observed due to suppressed interference, in the parallel case $g^{(2)}_{12}(\tau)$ starts from about 0.5 at small delays ($\tau < $100~$\mu$s) --which is close to the classical lower bound-- and increases to about 1 at $\tau > 1$~s, where the fluctuations are uncorrelated. The non-trivial behavior between these two limits is directly related to the complex structure of the noise spectrum shown~Fig.~\ref{Cryo_off} via the Wiener-Khinchin theorem. The binning of this measurement is finer than for the emitter case (Fig.~4 of the main text), with 10~bins per decade. Fig.~\ref{g2_laser}b shows the autocorrelation $g^{(2)}_{11}(\tau)$ for comparison. At $\tau > 1$~$\mu$s, the autocorrelation is symmetrical to the cross-correlation with respect to the horizontal axis $y = 1$. However, at shorter times, distortions due to dead time (at $\tau \leq 10^{-7}$~s) and afterpulse (at $\tau > 10^{-7}$~s) are visible. Although in theory $g^{(2)}_{11}$ and $g^{(2)}_{22}$ contain the same information as the cross-correlation $g^{(2)}_{12}$, these technical limitations prevent us to use these functions. In the experiments, we only use the cross-correlations functions, and we denote $g^{(2)}(\tau) = \frac{1}{2}\left( g^{(2)}_{12}(\tau) + g^{(2)}_{21}(\tau) \right)$.

\subsubsection{S1.3.2 Fluctuating source with finite bandwidth}

When the interferometer input is not monochromatic and originates from a real source of time-varying power spectrum $s(\omega,t)$, the interferometer outputs at a delay $\delta$, in the absence of phase fluctuations, are:
\[
I_{1,2}(t, \delta) = \dfrac{1}{2}I(t) \left( 1 \pm \Re \left[
g^{(1)}(\delta)
\right]
 \right)
=
\dfrac{1}{2}I(t) \left( 1 \pm \mathcal{F}_\omega \left[s(\omega,t)\right] \right)
\]
when the variation timescales of $s(\omega, t)$ are longer than both the lifetime (1.9~ns in our case) and the propagation time difference between the two arms of the interferometer ($\leq 1.3$~ns). The random phase $\phi(t)$ modifies this expression to
\[
I_{1,2}(t,\delta) = \dfrac{1}{2}I(t) \left( 1 \pm \cos[\phi(t)]\mathcal{F}_\omega \left[s(\omega,t)\right] \right)
\]

The second-order interferometric correlation (or cross-correlation) can be derived based on Eq.~\ref{crosscorr}, which yields:

\begin{equation}
g^{(2)}_{12}(\delta, \tau) = 
\dfrac{
\left\langle
I(t) I(t + \tau)
\right\rangle_t
}{
\left\langle
I(t) \right\rangle_t
\left\langle
I(t + \tau)
\right\rangle_t}
\left(
1 - C_\phi(\tau)
\left\langle
\mathcal{F}_\omega \left[s(\omega,t)\right]
\mathcal{F}_\omega \left[s(\omega,t + \tau)\right]
\right\rangle_t
\right)
\end{equation}

\begin{equation}
g^{(2)}_{12}(\delta, \tau) = 
g^{(2)}_\mathrm{HBT}(\tau)
\left(
1 - C_\phi(\tau)\mathcal{F}_\zeta[p(\zeta, \tau)])
\right)
\label{crosscorPCFS}
\end{equation}

where $g^{(2)}_\mathrm{HBT}(\tau)$ is the second-order correlation of the input signal as measured in the Hanbury Brown and Twiss configuration, and where we introduced the spectral correlation 
\[
p(\zeta, \tau) = \left\langle
\mathcal{F}_\omega \left[s(\omega,t)\right]
\mathcal{F}_\omega \left[s(\omega,t + \tau)\right] \right\rangle_t = \left\langle\mathcal{F}_\omega [s(\omega,t) * s(\omega,t + \tau)] \right\rangle_t
\]
where $*$ denotes the convolution product. The Fourier transform of this spectral correlation function is what we term the PCFS contrast of the emitter $C(\delta, \tau) = \mathcal{F}_\zeta[p(\zeta, \tau)]$. Similarly, 
\[g^{(2)}_{11}(\delta, \tau) = g^{(2)}_{22}(\delta, \tau) = g^{(2)}_\mathrm{HBT}(\tau)
\left(
1 + C_\phi(\tau)\mathcal{F}_\zeta[p(\zeta, \delta)])
\right)\]

To account for the case of a non-unity $g^{(2)}_\mathrm{HBT}(\tau)$, both cross- and autocorrelations can be combined to obtain the PCFS contrast~\cite{Coolen08}. Alternatively, the polarization can be set to be orthogonal, which provides $g^{(2)}_{ij,\perp}(\delta, \tau) = g^{(2)}_\mathrm{HBT}(\tau)$ $\forall (i,j)$. From Eq.~\ref{crosscorPCFS} we then obtain

\[
C(\delta, \tau) = \dfrac{1}{C_\phi(\tau)} \left(
1 - \dfrac{g^{(2)}_{\parallel}(\delta, \tau)}{g^{(2)}_{\perp}(\delta, \tau)}
\right) 
\] 

Finally, we note that $C(\delta = 0, \tau) = 1$, such that $C_\phi(\tau) = 1 - \dfrac{g^{(2)}_{\parallel}(\delta = 0, \tau)}{g^{(2)}_{\perp}(\delta = 0, \tau)}$. Therefore

\begin{equation}
C(\delta, \tau) = \dfrac{
1 - \dfrac{g^{(2)}_{\parallel}(\delta, \tau)}{g^{(2)}_{\perp}(\delta, \tau)}
}
{
1 - \dfrac{g^{(2)}_{\parallel}(\delta = 0, \tau)}{g^{(2)}_{\perp}(\delta = 0, \tau)}
}
\label{PCFS_general_formula}
\end{equation} 

This result shows that the PCFS contrast can be obtained from the interferometric correlation measured both in parallel and orthogonal configurations, without the need for any other characterization of the setup or the emitter. In particular, the statistics of the phase variations or fluctuations are absent from Eq.~\ref{PCFS_general_formula}, which is thus valid for both random fluctuations and deterministic variations of the interferometer phase difference.

In the absence of intensity fluctuations due to \textit{e.g.} blinking, $g^{(2)}_\mathrm{HBT}(\tau) = 1$, which is verified in our experiment at the experimentally accessible values of $\tau$. In this case, Eq.~\ref{PCFS_general_formula} further simplifies to

\begin{equation}
C(\delta, \tau) = \dfrac{
1 - g^{(2)}_{\parallel}(\delta, \tau)}
{
1 - g^{(2)}_{\parallel}(\delta = 0, \tau)
}
\label{PCFS_g2_1}
\end{equation} 
\textit{i.e.} at a given $\tau$, the PCFS contrast at optical delay $\delta$ is the degree of antibunching normalized by its value at zero optical delay.

Finally, we note that, thanks to the self-normalization, the expressions of $C(\delta, \tau)$ in equations~\ref{PCFS_general_formula} and~\ref{PCFS_g2_1} are robust to uncorrelated background noise (such as dark count or ambiant light), which decreases the degree of antibunching independently of $\tau$ and $\delta$.

\section{S2. Spectral correlation function, spectral diffusion kernel and effective spectrum}

In this section, we derive the relation between the spectral correlation function $p(\zeta, \tau)$, to which the PCFS technique gives access, and the SD kernel $P(\omega,\zeta; \tau)$ used to describe microscopic models. This relation makes it possible to infer the expected results of PCFS measurements for any given SD process, either analytically or using Monte Carlo simulations in the general case. In addition, we also provide the relation between $p(\zeta, \tau)$ and the effective time-dependent spectrum $S_\mathrm{eff}(\omega, \tau)$, which can be used to derive the time-dependent coherence time and linewidth.

The PCFS technique provides access to the spectral correlation function

\begin{equation}
p(\zeta, \tau) = \left\langle \int d\omega 
s(\omega, t) s(\omega + \zeta, t + \tau)
\right\rangle_t
\label{spec_correl}
\end{equation}
which is the time-averaged correlation of instantaneous spectra $s(\omega,t)$ and $s(\omega,t + \tau)$ separated by $\tau$. Denoting by $\omega_c(t)$ the instantaneous center frequency, and by $S_\mathrm{hom} (\omega) $ the homogeneous lineshape (taken to be centered at zero), we have $s(\omega,t) = S_\mathrm{hom} (\omega) * \delta(\omega - \omega_c(t))$. Denoting by $\star$ the correlation operator, \textit{i.e.} $(f \star g)(\zeta) = \int d\omega f(\omega) g^*(\omega + \zeta)$, the spectral correlation can be rewritten
\begin{align*}
p(\zeta, \tau) &= 
\left\langle 
s(\omega, t) \star s(\omega + \zeta, t + \tau)
\right\rangle
\\
&= 
\left\langle 
(S_\mathrm{hom}(\omega) * \delta(\omega - \omega_c(t))) \star (S_\mathrm{hom}(\omega) * \delta(\omega - \omega_c(t + \tau)))
\right\rangle
\\
&= \underbrace{(S_\mathrm{hom} \star S_\mathrm{hom} )}_{p_\mathrm{hom}(\zeta)}
*
\underbrace{
\left\langle 
\delta(\omega_c(t + \tau) - \omega_c(t))
\right\rangle
}_{p_\mathrm{inhom}(\zeta, \tau)}
\end{align*}

$p(\zeta, \tau)$ is therefore the convolution product of two terms. The first is the autocorrelation of the (deterministic) homogeneous spectral lineshape $p_\mathrm{hom}(\zeta) = (S_\mathrm{hom} \star S_\mathrm{hom} )(\zeta)$. The second one, $p_\mathrm{inhom}(\zeta, \tau)$, depends on the stochastic process governing $\omega_c(t)$, and can be explicited from the spectral diffusion kernel $P(\omega,\zeta; \tau) = P(\omega + \zeta, t + \tau|\omega, t)$ using Bayes theorem, which yields:

\begin{align}
p_\mathrm{inhom}(\zeta, \tau) &= \int d\omega P(\omega, t) P(\omega + \zeta, t + \tau|\omega, t) \nonumber
\\
&= \int d\omega S_\mathrm{inhom}^\infty(\omega) P(\omega,\zeta; \tau)
\label{p_kernel}
\end{align}

Contrarily to $p_\mathrm{hom}(\zeta)$, the expression of $p_\mathrm{inhom}(\zeta, \tau)$ is not an explicit autocorrelation function. However, if the SD process is time-reversible, $p_\mathrm{inhom}(\zeta, \tau)$ can always be written as such, \textit{i.e.} $\exists S_\mathrm{inhom}(\omega, \tau)$ such that $p_\mathrm{inhom}(\zeta, \tau) = S_\mathrm{inhom} (\omega, \tau) \star S_\mathrm{inhom}(\omega, \tau)$.

In the general case, $S_\mathrm{inhom}(\omega, \tau)$ can be expressed using the Wiener–Khinchin theorem as 

\begin{equation}
S_\mathrm{inhom}(\omega, \tau) = FT^{-1}
\left[
\sqrt{FT
\left[
p_\mathrm{inhom}(\zeta, \tau)
\right]
}
\right]
\label{p_s}
\end{equation}

This yields the following simple expression:
\begin{equation*}
p(\zeta, \tau) = p_\mathrm{hom}(\zeta) * p_\mathrm{inhom}(\zeta, \tau) = (S_\mathrm{hom} \star S_\mathrm{hom} ) * (S_\mathrm{inhom} \star S_\mathrm{inhom} )
\label{p_FT}
\end{equation*}

In analogy to the steady-state spectrum $S^\infty(\omega) = S_\mathrm{hom}(\omega) * S_\mathrm{inhom}^\infty(\omega)$, we can define the time-dependent effective spectrum $S_\mathrm{eff}(\omega, \tau)$ as the convolution product of the homogeneous lineshape by the time-dependent inhomogeneous envelope: $S_\mathrm{eff}(\omega, \tau) = S_\mathrm{hom}(\omega) * S_\mathrm{inhom}(\omega, \tau)$. The spectral function simply writes 

\begin{equation}
p(\zeta, \tau) = S_\mathrm{eff} \star S_\mathrm{eff} = \int d\omega S_\mathrm{eff}(\omega, \tau) S_\mathrm{eff}(\omega + \zeta, \tau)
\label{p_eff_spectrum}
\end{equation}

This expression generalizes the short-time limit, where $p(\zeta, \tau = 0) = S_\mathrm{hom} \star S_\mathrm{hom}$ and the long-time limit $p(\zeta, \tau \rightarrow \infty) = S^\infty \star S^\infty$ to any finite delay $\tau$. In other words, the spectral correlation function $p(\zeta)$ can be seen as the autocorrelation of a time-dependent, effective spectrum $S_\mathrm{eff}(\omega, \tau)$.

Comparing Eqs~\ref{spec_correl} and \ref{p_eff_spectrum} leads to the interpretation of $S_\mathrm{eff}(\omega, \tau)$ as a fixed, deterministic spectrum whose autocorrelation identifies with the time-averaged correlation of two homogeneous but randomly displaced spectra separated by~$\tau$.

The notable limits of $S_\mathrm{inhom}(\omega,\tau)$ are $S_\mathrm{inhom}(\omega,\tau \rightarrow 0) = \delta(\omega)$ and $S_\mathrm{inhom}(\omega,\tau \rightarrow \infty) = S^\infty(\omega)$, so that the (total) effective spectrum $S_\mathrm{eff}(\omega, \tau)$ evolves from the homogeneous line $S_\mathrm{hom}(\omega)$ at $\tau = 0$ to the full homogeneous spectrum $S^\infty(\omega)$ at $\tau \rightarrow \infty$. At intermediate scales, its shape and width are dictated by the SD kernel, so that it carries the relevant information about SD. In particular, its linewidth allows to define a time-dependent coherence time (see Section S3). Finally, the PCFS contrast $C(\delta, \tau)$ is simply given by the squared modulus of the Fourier transform of $S_\mathrm{eff}(\omega, \tau)$.


\section{S3. Effective SD spectrum of a few processes}

In this section, we explicitly derive $p(\zeta, \tau)$ and $S(\omega, \tau)$ in the case of two common SD mechanisms.

\subsection{S3.1 Continuous Gaussian diffusion -- Ornstein Uhlenbeck process}

In the Ornstein-Uhlenbeck (OU) process associated with the inhomogeneous distribution $\Delta\omega_\mathrm{inhom}^\infty = 2 \sqrt{2 \ln 2} \Sigma^\infty$, the SD kernel writes~\cite{Delteil24, Gaignard25}:

\begin{equation*} 
P(\omega, \zeta ; \tau) = \frac{1}{\sqrt{2 \pi} \Sigma^\infty \sqrt{1 - e^{-2\tau/\tau_\mathrm{SD}}}}
\exp\left[-
\dfrac{\left( \omega \left( 1 - e^{-\tau/\tau_\mathrm{SD}} \right) + \zeta \right)^2}
{2 \Sigma_\infty^2 (1 - e^{-2\tau/\tau_\mathrm{SD}})} 
 \right]
\label{OU_kernel}
\end{equation*}
where the center of the inhomogeneous spectrum has been taken to be at zero for simplicity. Injecting this expression in Eq.~\ref{p_kernel} and using $S_\mathrm{inhom}^\infty(\omega') = \lim_{\tau \rightarrow \infty} P(\omega, \zeta ; \tau)$  leads to 

\begin{equation*} 
p_\mathrm{inhom}(\zeta, \tau) = \frac{1}{2\sqrt{ \pi} \Sigma(\tau)} \exp{\left(- \dfrac{\zeta^2}{4 \Sigma(\tau)^2}  \right)}
\label{OU_p}
\end{equation*}
with $\Sigma(\tau) = \Sigma^\infty \sqrt{1 - e^{-\tau/\tau_\mathrm{SD}}}$

The time-dependent inhomogeneous envelope $S_\mathrm{inhom}(\omega, \tau)$ should have, by definition, its autocorrelation function equal to $p_\mathrm{inhom}(\zeta, \tau)$. This directly yields

\begin{equation*} 
S_\mathrm{inhom}(\omega, \tau) = \frac{1}{\sqrt{ 2\pi} \Sigma(\tau)} \exp{\left(- \dfrac{\omega^2}{2 \Sigma(\tau)^2}  \right)}
\label{OU_s}
\end{equation*}
\textit{i.e.} $S_\mathrm{inhom}(\omega, \tau)$ is a Gaussian function which, at short times, tends to $\delta(\omega)$ and at long times identifies with the inhomogeneous spectrum $S_\mathrm{inhom}^\infty(\omega)$. The choice of the distribution center is arbitrary (since $S_\mathrm{inhom}$ is defined by its autocorrelation) and can be taken to match that of the inhomogeneous distribution. This expression can be generalized to arbitrary broadening dynamics described by $\Sigma(\tau)$. The effective spectrum $S_\mathrm{eff}(\omega, \tau)$ is obtained by convolving $S_\mathrm{inhom}(\omega, \tau)$ with a Lorentzian homogeneous lineshape $S_\mathrm{hom}(\omega)$, which provides a Voigt profile. Taking the squared modulus of the Fourier transform, we obtain the expression given in Eq.~4 of the main text to obtain the associated PCFS contrast, which reads as the product of a Gaussian and an exponential decay.

\subsection{S3.2 Discrete Gaussian random jumps}

In the case of a discrete process consisting in random jumps within a fixed envelope $S_\mathrm{inhom}^\infty(\omega)$, the SD kernel can be expressed as

\begin{equation*} 
P(\omega, \zeta ; \tau) = a(\tau) \delta(\zeta)+\left( 1 - a(\tau) \right) S_\mathrm{inhom}^\infty(\omega + \zeta)
\label{GRJ_kernel}
\end{equation*}
where the relative weight of non-diffused spectrum $a(\tau)$ describes the SD dynamics. Indeed, given a spectral position $\omega$ of the emitter at time $t$, there is a probability $a(\tau)$ that there has been no jump between $t$ and $t + \tau$ -- yielding a $\delta(\zeta)$ contribution -- and a probability $ 1 - a(\tau)$ that at least one jump has occurred, placing the emitter in a new, uncorrelated spectral position within the inhomogeneous envelope $S_\mathrm{inhom}^\infty(\omega + \zeta)$, which describes the probability density function of the updated spectral position $\omega + \zeta$. In the case of a Poisson process where the jump times are uncorrelated, $a(\tau) = e^{-\tau/\tau_\mathrm{SD}}$.

The spectral correlation is then 
\begin{align*} 
p_\mathrm{inhom}(\zeta, \tau) = a(\tau) \delta(\zeta)+\left( 1 - a(\tau) \right) 
\left(
S_\mathrm{inhom}^\infty \star S_\mathrm{inhom}^\infty
\right)
(\zeta) 
\label{GRJ_p}
\end{align*}

This expression matches the one postulated in the Supplemental Material of reference~\cite{Utzat19}, and readily provides Eq.~5 of the main text. While $S_\mathrm{inhom}(\omega, \tau)$ does not have an exact analytical expression~\cite{footnote}, an excellent approximation can be derived as

\begin{align*} 
S_\mathrm{inhom}(\omega, \tau)  = a(\tau) \delta(\omega)+\left( 1 - a(\tau) \right) 
\left(
\dfrac{1}{\sqrt{2 \pi} \Sigma_\mathrm{eff}} \exp \left[
- \dfrac{\omega^2}{2 \Sigma_\mathrm{eff}^2}
\right]
\right)
\end{align*}

with $\Sigma_\mathrm{eff} = \Sigma^\infty \sqrt{ \dfrac{1}{1 + 3 a(\tau)} }$. This effective inhomogeneous spectrum is the sum of a Dirac distribution (undiffused case) and a Gaussian distribution (diffused case) and identifies with the time-averaged inhomogeneous distribution at long times.

\end{document}